\documentclass{aa}
\usepackage{txfonts}
\usepackage{graphicx}
\usepackage{natbib}

\begin{document}
%\title{Improved cool white dwarfs atmosphere models in the analysis of the white dwarf cooling sequence of NGC 6397} 
\title{White dwarf atmosphere models with Ly-$\alpha$ opacity in the analysis of the white dwarf cooling sequence of NGC 6397} 

%\subtitle{II. An example text with infinitesimal
%  scientific value\\ 
%  whose title and subtitle may also be split} 

\author{Piotr M. Kowalski}
%  \and Apostolos Hadjidimios\inst{2} 
%  \thanks{\emph{Present address:} 
%    Department of Computer Science, Purdue University, 
%    West Lafayette, IN 47907, USA} 
%     \and Robert J. Plemmons\inst{3}} 

\offprints{Piotr Kowalski, \email{ piotr.kowalski@theochem.ruhr-uni-bochum.de or pmkowalski@aol.com 
This email address is being protected from spam bots, you need Javascript enabled to view it }}

\institute{Lehrstuhl f\"ur Theoretische Chemie, Ruhr-Universit\"at, 44780 Bochum, Germany
%  \and Department of Mathematics, University of Ioannina, 
%  GR-45 1210, Ioannina, Greece 
%  \and Department of Computer Science and Mathematics, 
% North Carolina State University, Raleigh, NC 27695-8205, USA
} 

%\date{2 April 2007}

\abstract{} 
{
We discuss the importance of pure hydrogen white dwarf atmosphere models with Ly-$\rm \alpha$ far red wing 
opacity in the analysis of the white dwarf cooling sequence of the
globular cluster NGC 6397.}
{Our recently improved atmosphere models account for the previously missing opacity from the Ly-$\rm \alpha$ hydrogen line broadened by collisions 
of the absorbing hydrogen atoms with molecular and atomic hydrogen. These models are the first 
that well reproduce the UV colors and spectral energy distributions of cool white dwarfs with $T_{\rm eff}<6000 \rm \, K$ observed in the Galactic Disk. 
}
{Fitting the observed $F814W$ magnitude and $F606W-F814W$ color we obtained a value for the true distance modulus, $\mu=12.00 \pm 0.02$, that is in agreement with recent analyses. 
%However, 
%with our new models the resulting reddening toward the cluster $E(F606-F814)=0.12 \pm 0.02$ is slightly lower. 
We show that the stars at the end of the cooling sequence appear to be $\rm \sim 160 \, K$ cooler
when models that account for Ly-$\rm \alpha$ opacity are used. 
%in the analysis.
This indicates that the age of NGC 6397 derived from the white dwarf cooling sequence using atmosphere models that do not include the 
correct Ly-$\alpha$ opacity is underestimated by $\sim \, 0.5$ Gyr. 
} 
{Our analysis shows that it is essential to use white dwarf atmosphere models with Ly-$\rm \alpha$ opacity
for precise dating of old stellar populations from white dwarf cooling sequences. 
}


\date{Received 31 May 2007 / Accepted 27 June 2007}

\authorrunning{Kowalski}
\titlerunning{Cool WDs in NGC 6397}

\maketitle

\section{Introduction}
Globular clusters are the oldest known stellar systems in our Galaxy. 
%They are almost as old as 
%the Universe itself. 
They contain a large population of very cool and very old white dwarfs, remnants of stars that formed and died 
through the history of a cluster. Such populations have been investigated in the two nearest globular clusters: M4 \citep{HN02,RI04,HN04} and NGC 6397 \citep{HN07,RI06}. 
Recent data on NGC 6397 show a clear white dwarf cooling sequence and a low luminosity cut off related to the finite age of the cluster.
On the basis of this sequence, \citet{HN07} derived the true distance modulus $\mu_{0}=12.02\pm0.06$, the reddening toward NGC 6397 $E(F606W-F814)=0.20 \pm 0.03$, 
and its age $T_{c}=11.47 \pm 0.47$ Gyr. The detailed and extended analysis presented in that publication gives the first precise and independent 
dating of a globular cluster based on its white dwarf cooling sequence.

\begin{figure}
\resizebox{\hsize}{!}{\rotatebox{270}{\includegraphics{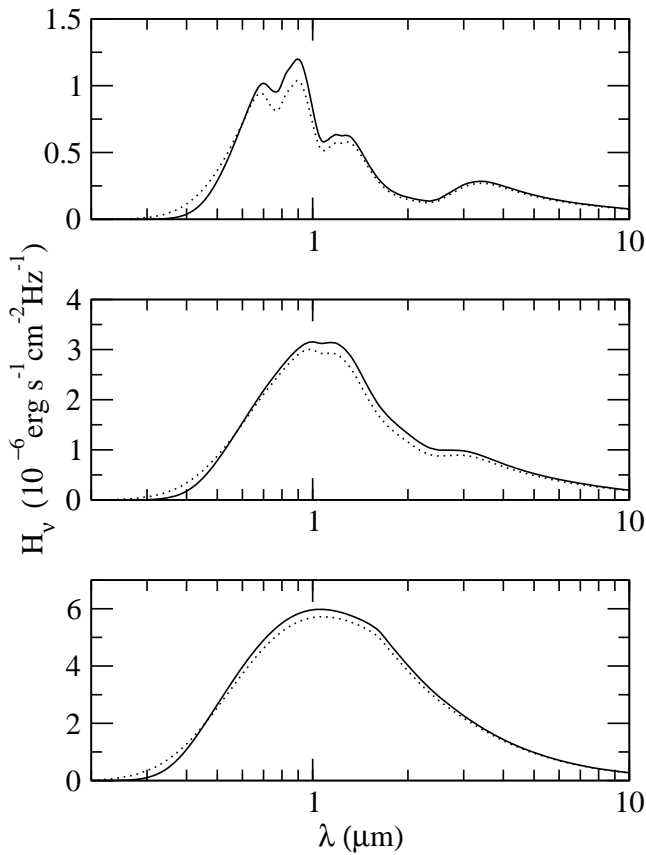}}}
\caption{Synthetic spectra of pure hydrogen atmosphere models with 
$T_{\rm eff}=3000 \rm
\, K$, $4000 \rm \, K$, and $5000 \, \rm K$ (from top to bottom), and fixed 
gravity $log \, g \rm = 8 \, (cgs)$
calculated with (solid) and without (dotted) Ly-$\rm \alpha$ far red wing opacity.
The dotted lines mimic the models of \citet{BSW95} and \citet{SJ} used by \citet{HN07} and \citet{RI06} in their analysis of the white dwarf cooling
sequence of NGC 6397.
\label{F1}}
\end{figure}

%To achieve a full realization of white dwarf cosmochronology the 
Reliable atmosphere models need to be used
in such analyses \citep{FN00,HN99}. Models currently in use (eg. \citet{BLR,SJ,HN98,BSW95}), although valuable in the analysis of cool white dwarfs (eg. \citet{BLR}), 
have a well known shortcoming. They cannot correctly reproduce the UV flux ($\rm \lambda<5000 \, \AA$) of white dwarfs with $T_{\rm eff}< \rm 6000 \, K$ \citep{Bergeron01,BLR,Bergeron97}. 
This problem has been solved recently by introducing the far red wing profile of hydrogen Ly-$\alpha$ into the modeling. 
The line profile is formed by collisions between absorbing hydrogen atoms and the atomic and molecular hydrogen, and when introduced into the modeling 
it allows for an accurate reproduction of the spectral energy distribution of cool white dwarfs
at all wavelengths \citep{KS06}. The synthetic spectra resulting from our improved pure hydrogen atmosphere models deviate significantly from the previous ones for wavelengths shorter than $1 \, \rm \mu m$ 
(Fig. \ref{F1}). This indicates that the Ly-$\alpha$ opacity also affects the flux in filters $F606W$ and $F814W$ used to obtain the photometric data on the white dwarf cooling sequence of NGC 6397. 

In this paper we discuss the importance of our new white dwarf atmosphere models with Ly-$\alpha$ opacity in the analysis of white dwarf cooling sequences. 
Fitting the theoretical to observed white dwarf cooling sequence of NGC 6397 we derived the true distance modulus to the cluster. We show that its value is consistent with that 
obtained with old models. 
%On the other hand the reddening toward the cluster derived here is slightly lower. 
We also predict that the coolest white dwarfs observed in NGC 6397 
are cooler, when the atmosphere models with Ly-$\alpha$ opacity are used in the analysis. This implies that the cluster age derived from the white dwarf cooling sequence by 
\cite{HN07} is probably too low. Comparing the temperature of the white dwarfs at the cut off of the white dwarf cooling sequence in NGC 6397 and the galactic disk 
we show that the local population of disk white dwarfs is much younger than that observed in the globular cluster.  

%In the next section we discuss the impact of the Ly-$\alpha$ opacity for the photometry measured in $F606W$ and $F814W$ filters.  
%Our analysis of white dwarf cooling sequence of NGC 6397 follows in section 3. In section 4 we present the discussion of the derived age of the globular cluster.
%Conclussions follow in section 5.

\section{Cool white dwarf atmosphere models}
\citet{HN07} and \citet{RI06} in their analysis of the white dwarf cooling sequence of NGC 6397 used the atmosphere models of \citet{BLR}.
%of \citet{BSW95} and \citet{SJ}.
However, there is a well known shortcoming of these models, as they overestimate the flux at short wavelengths ($\lambda<5000 \, \AA$, see \citet{Bergeron97,Bergeron01}).
We have corrected this deficiency by introducing the opacity of the far red wing of the Ly-$\rm \alpha$ broadened by collisions 
of the absorbing hydrogen atoms with atomic and molecular hydrogen \citep{KS06}. Our new models allow for fits to the entire spectral energy distribution of white dwarfs observed in our Galaxy,
including the fit to the near-UV HST spectrum of the white dwarf star BMP 4729 \citep{KS06,Koester00}, and the reproduction of the observed white dwarf color sequences below $T_{\rm eff} \rm \sim 6000 \, K$ \citep{KS06}.  

\begin{figure}
\resizebox{\hsize}{!}{\rotatebox{270}{\includegraphics{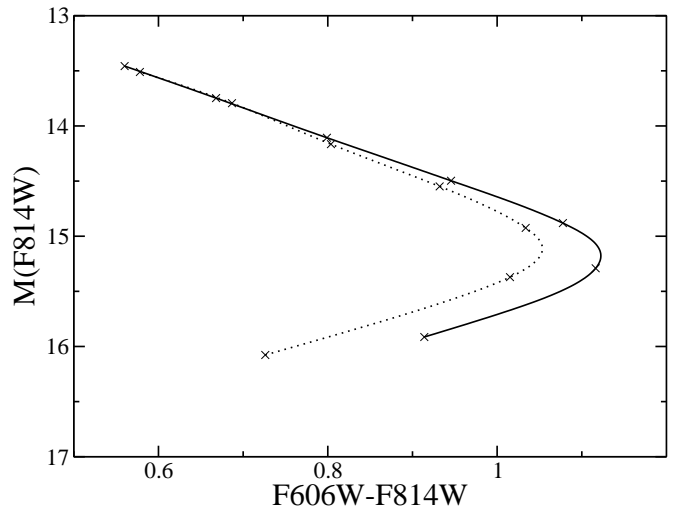}}}
\caption{Color-magnitude diagram
of the white dwarf cooling sequence. The mass of the star is $0.5 \, M_{\sun}$, and the effective temperature
runs from $\rm 6000 \, K$ (top) to $\rm 3000 \, K$ (bottom). 
Crosses indicate the $T_{\rm eff}$ with $\rm 500 \, K$ decrements along the curves.
Predictions of both models with (solid) and without (dotted) Ly-$\rm \alpha$ opacity
are shown. On the vertical axis the absolute magnitude is indicated.
%The dotted lines mimics the models of \citet{BSW95}.
\label{F2}}
\end{figure}

In Fig. \ref{F1} we present the impact of the Ly-$\rm \alpha$ opacity on the synthetic spectra of cool white dwarfs. 
The flux suppressed by the Ly-$\rm \alpha$ opacity at short wavelengths is redistributed towards longer wavelengths (well beyond $\lambda \rm \sim 1 \, \mu m$).
This causes a non negligible increase of the flux at visible and near infrared wavelengths. As a result, the 
synthetic photometry in the $F606W$ and $F814W$ filters is also affected. This is shown in figure \ref{F2} where we present the color-magnitude diagram for
a cooling sequence of a white dwarf of mass $0.5 \, M_{\sun}$ computed using models 
with and without the Ly-$\rm \alpha$ opacity. Both curves deviate significantly from each other  
for $T_{\rm eff}\rm <5000 \, K$. 
As Ly-$\rm \alpha$ opacity is stronger at short wavelengths, the effect is stronger in the $F606W$ filter (larger horizontal shift in Fig. 2). 
The new models predict a turn towards the blue, caused by CIA opacity from molecular hydrogen \citep{SJ,HN98,BSW95}, at lower $T_{\rm eff}$. 
This should have an impact on the astronomical parameters of NGC 6397 derived from the analysis 
of its white dwarf cooling sequence.
% $T_{\rm eff}$ and the ages of the coolest white dwarfs observed in this globular cluster.   

\section{Analysis of the data}

\subsection{The fit to the white dwarf cooling sequence}

\begin{figure}
\resizebox{\hsize}{!}{\rotatebox{270}{\includegraphics{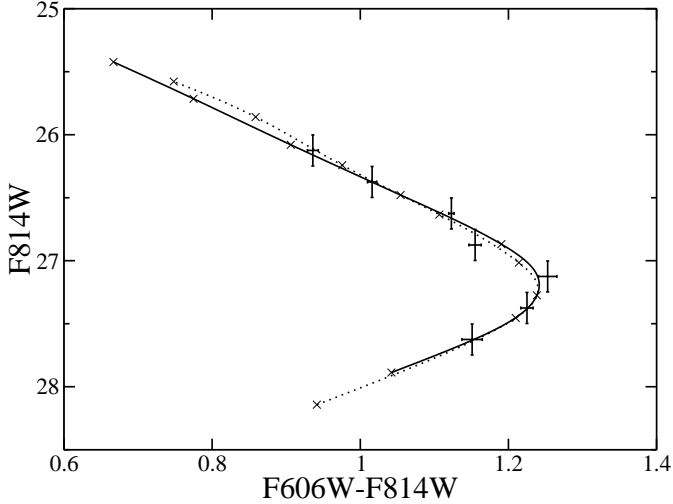}}}
\caption{Color-magnitude diagram
of the white dwarf cooling sequence observed in NGC 6397. The lines represent the 
fits to data using modeled cooling sequences for a white dwarf of mass $0.5 \, M_{\sun}$.
The effective temperature runs from $\rm 6000 \, K$ (top) to $\rm 3000 \, K$ (bottom). 
Small crosses indicate the $T_{\rm eff}$ with $\rm 500 \, K$ decrements along the curves.
Models with (solid) and without (dotted) Ly-$\rm \alpha$ opacity are shown. 
%The dotted lines mimics the models of \citet{BSW95}.
\label{F3}}
\end{figure}

To investigate the importance of pure hydrogen white dwarf atmosphere models with Ly-$\alpha$ opacity in the analysis of the white dwarf cooling sequence of NGC 6397 
we performed a least square 2D fit in F814W and F606W-F814W to the photometric data of \citet{HN07,RI06}; their Fig. 6 and 4 respectively. 
The fitting parameters were the reddening and the true distance modulus.
The fit was performed using two grids of atmosphere models: 
those with Ly-$\rm \alpha$ opacity and those without. The second set mimics the models used in the analysis of the cluster white dwarf cooling sequence conducted 
by \citet{HN07} and \citet{RI06}. Following those authors, in our derivation of the theoretical cooling sequences we assumed the mass of the white dwarf to be $0.5 \, M_{\sun}$.
Our fits to the data are given in Fig. \ref{F3}. We obtained perfect fits with both theoretical cooling sequences, as is also the case 
in \citet{HN07} and \citet{RI06}. In addition to that work we show, for the first time, the $T_{\rm eff}$ of cool white dwarfs in the globular cluster.
Because of the difference in localization of the turn off visible in both theoretical sequences (Fig. \ref{F2}), both fits give different estimations 
of $T_{\rm eff}$ for the cool end of the observed sequence localized at F814W$\sim \, 27.6$ \citep{HN07}. The effective temperature estimated on the basis of atmosphere 
models that account for Ly-$\rm \alpha$ opacity is significantly smaller, indicating that the stars at the cut off 
of the observed white dwarf cooling sequence are cooler and older. We will return to this problem in the next section.

\subsection{Reddening and the true distance modulus}
As indicated in Figure \ref{F2} there is a significant variation in the modeled maximum value of the $F606W-F814W$ color 
for a white dwarf of a given mass.
%that a white dwarf of a given mass can posses. 
In the present case, both model curves deviates by $\sim 0.1$ magnitude, which has a direct impact on the derived value for the reddening towards NGC 6397.
%in the considered filters. 
With models that do not account for the Ly-$\rm \alpha$ opacity 
we obtained $E(F606W-F814W)=0.18 \pm 0.2$ with a good agreement with the value 
%$0.17 - 0.22$ 
$0.16$
obtained by \citet{HN07}. Models with the Ly-$\alpha$ give a lower value
of $0.12 \pm 0.02$. This result, when converted from the ACS photometric system to the ground-based colors using the prescription of \citet{SIR05} gives $E(B-V)=0.13 \pm 0.03$. 
This value is lower than $E(B-V)=0.18$ obtained by \citet{GR03} using a fit to the Main Sequence stars. 
However, the reddening obtained by fitting the $M=0.5M_{\sun}$ theoretical cooling sequence to the photometric data is expected to be underestimated. This is because 
of the variation in the average mass of a white dwarf at the end of the cooling sequence. This is discussed in detail by \citet{HN07}. The final reddening derived
in that paper is $E(F606W-F814W)=0.20 \pm 0.3$. 
This is $\sim 0.04$ dex larger than the value they obtained using the $M=0.5M_{\sun}$ theoretical cooling sequence. 
Applying such a shift to our result, we obtain $E(F606W-F814W)=0.16 \pm 0.02$ and $E(B-V)=0.17 \pm 0.03$ respectively.
This value is in good agreement with the result of \citet{GR03}.         
%
%Our fits do not perfectly 
%reproduce the data point of the largest $F606W-F814W$ value. This data point is outside the theoretical sequences by $\sim 0.01$ dex. It is therefore possible that
%more restricted fit would push the theoretical prediction further to the left, increasing the derived reddening and the agreement with result of \citet{GR03}. 
%Because of the Ly-$\rm \alpha$ opacity 
%the cool white dwarfs posses highly suppressed flux in the $B$ filter, which is not true in the case of other stars. 
%Therefore it is also possible that there may be a systematic error related to this effect, when the transformation from $E(F606W-F814W)$ to $E(B-V)$ of \citet{SIR05} is used.
%This would also lead to an underestimation of the $E(B-V)$.  

For the true distance modulus, in both fits we obtained a consistent value of $12.00 \pm 0.02$, which agrees with the value obtained by \citet{HN07}, $\mu = 12.02 \pm 0.06$. 
Our estimate of the true distance modulus was obtained assuming a single white dwarf mass of $M=0.5M_{\sun}$ and the relation between reddening in both filters $A_{814}=0.65A_{606}$ \citep{SIR05}.
When we use the $M=0.6M_{\sun}$ mass star instead, the resulting distance modulus is lower, $11.85\pm0.02$.

\begin{table}[t]
\caption{Effective temperature and ages of white dwarfs at the end of the cooling sequence of NGC 6397.} 
%\tablewidth{0pt}
%\tablehead{& \colhead{WD}
\centering
\begin{tabular}{c c c c c}
\hline\hline
mass ($M_{\rm \sun}$) & $T_{\rm eff} \, \rm (K)$ & age (Gyr) & $T_{\rm eff} \, \rm (K)$ & age (Gyr) \\
\hline
& \multicolumn{2}{c}{models with Ly-$\rm \alpha$} & \multicolumn{2}{c}{models without Ly-$\rm \alpha$} \\
0.5 & 3180 $\pm$ 80 & 9.8 $\pm$ 0.4 & 3340 $\pm$ 80 & 9.2 $\pm$ 0.3  \\
0.6 & 3210 $\pm$ 80 & 11.1 $\pm$ 0.3 & 3380 $\pm$ 90 & 10.6 $\pm$ 0.3  \\

\hline
\label{T3}
\end{tabular}
\end{table}

\begin{table}[b]
\caption{Effective temperature of the white dwarfs at the end of the cooling sequence
observed in the Galactic Disk.} 
%\tablewidth{0pt}
%\tablehead{& \colhead{WD}
\centering
\begin{tabular}{c c c}
\hline\hline
color & value & $T_{\rm eff}$ (K) \\
\hline
B-V & 1.4 $\pm$ 0.1 & 3750 $\pm$ 250 \\
V-I & 1.4 $\pm$ 0.1 & 3800 $\pm$ 300 \\
u-g & 2.6 $\pm$ 0.1 & 3750 $\pm$ 150 \\
g-z & 1.9 $\pm$ 0.1 & 3850 $\pm$ 150 \\ 

\hline
\label{T2}
\end{tabular}
\end{table}

\section{
$T_{\rm eff}$ at the edge of the white dwarf cooling sequence and the age of NGC 6397}
The proper derivation of the age of NGC 6397 from its white dwarf cooling sequence is a difficult task and is beyond the scope of this paper. 
However, as use of the new atmosphere models shifts the effective temperature of the coolest white dwarfs by a few hundred degrees (Fig. \ref{F2}), 
it is important to show how large a bias this effect would produce on the derived age of NGC 6397.
To check this we conducted a simple exercise, similar to that of \citet{HN07}, appendix A. Following the authors we assumed that the observed cut off of the white dwarf cooling
sequence takes place at $F814W=27.6 \pm 1$. Then assuming the white dwarf masses to be $0.5 \, M_{\sun}$ 
and $0.6 \, M_{\sun}$ respectively we derived the $T_{\rm eff}$ and, using the same white dwarf cooling models of \citet{RI00} as \citet{HN07}, 
we derive the cooling times of the white dwarfs at the cooling sequence cut off. The results are summarized in Table \ref{T3}. The $T_{\rm eff}$ and the ages derived using atmosphere models 
with and without Ly-$\alpha$ opacity differ by $\sim 160 \, \rm K$ and $\sim \, 0.5 \, \rm Gyr$ respectively, and these differences are insensitive to the assumed white dwarf mass. 
This suggests that the age of the globular cluster derived in \citet{HN07} can be $\sim 0.5$ Gyr too low because 
%of the usage 
of the white dwarf atmosphere models that 
do not account for Ly-$\rm \alpha$ red wing opacity.

\begin{figure}[t]
\resizebox{\hsize}{!}{\rotatebox{270}{\includegraphics{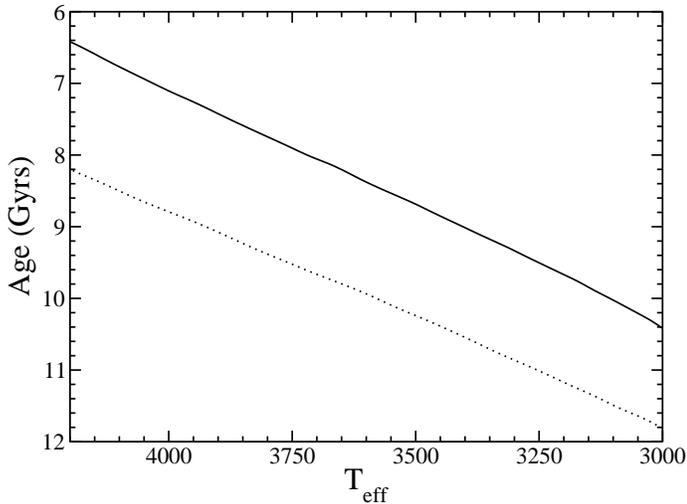}}}
\caption{White dwarf cooling times vs. 
$T_{\rm eff}$ for a white dwarf of mass 
$0.5 \, M_{\sun}$ (solid line) and $0.6 \, M_{\sun}$ (dotted line). Models of \citet{RI00}. \label{F4}}
\end{figure}

We also conducted the same simple analysis for the white dwarf cooling sequence observed in the Milky Way obtained from three samples of white dwarfs \citep{BLR,Kilic06a,Har06}. 
To do this we used published color sequences of \citet{KS06}. The obtained value of colors and $T_{\rm eff}$ at the cut off  
are given in Table \ref{T2}. Our analysis shows that the coolest white dwarfs observed in the Galactic Disk have $T_{\rm eff}\rm \sim 3800 \, K$. This indicates that the coolest stars in 
the local population of white dwarfs are warmer by $\rm \sim 600 \, K$ than the coolest white dwarfs observed in NGC 6397. According to white dwarf cooling models of \citet{RI00} 
such a difference in effective temperature results in $\sim \, 2$ Gyr difference between the age of the white dwarfs in NGC 6397 and the local white dwarf population (Fig. \ref{F4}). 
This is consistent with the result of \citet{FN00}, who, based on the analysis of the white dwarf luminosity functions, concluded that the Galactic Halo is $2-3$ Gyr older than the Galactic Disk.

\section{Conclusions}
Observation of the white dwarf cooling sequences in the globular clusters opens the possibility to date them more precisely.
However, there are many factors that affect the models applied to white dwarf cosmochronology. Here we have discussed the impact of white dwarf atmosphere models on the
derivation of the basic astronomical properties of NGC 6397 from the white dwarf cooling sequence itself. Our most recent set of pure hydrogen models, 
that corrects for the 
%long existed 
flux discrepancy at short wavelengths, results in lower reddening toward the cluster,
and older derived ages. By deriving the effective temperature of white dwarfs at the end of the white dwarf cooling sequences of NGC 6397 and the Galactic Disk
we have also shown that the coolest white dwarfs observed in the globular cluster are $\sim 600 \, \rm K$ cooler, and therefore $\sim 2 \rm \, Gyr$ older.  
More sophisticated analysis that includes the construction of the white dwarf luminosity function 
%using the new white dwarf atmosphere models, that is beyond our capabilities, 
is necessary for correct derivation of the cluster age from the white dwarf cooling sequence; however, our analysis shows that atmosphere models with Ly-$\rm \alpha$ opacity 
are necessary to reliably estimate the ages of old stellar populations 
from the white dwarf cooling sequences \footnote{Colors available upon request.}.

\begin{acknowledgements}
I thank Didier Saumon for useful comments on this manuscript and the theoretical chemistry group at Ruhr University in Bochum for hospitality and support.
\end{acknowledgements}


\begin{thebibliography}{}
\bibitem[Bergeron, 2001]{Bergeron01} Bergeron, P. 2001, \apj, 558, 369
\bibitem[Bergeron et al., 2001]{BLR} Bergeron, P., Leggett, S. K., \& Ruiz, M. T. 2001, \apjs, 133, 433
\bibitem[Bergeron et al., 1997]{Bergeron97} Bergeron, P., Ruiz, M. T., \& Leggett, S. K. 1997, \apj, 108, 339
\bibitem[Bergeron et al., 1995]{BSW95} Bergeron, P., Saumon, D., \& Wesemael, F. 1995, \apj, 443, 764 
\bibitem[Fontaine et al., 2001]{FN00} Fontaine, G., Brassard, P. \& Bergeron, P. 2001, PASP, 113, 409 
\bibitem[Gratton et al., 2003]{GR03} Gratton, R. G., Bragaglia, A., Carretta, E., Clementini, G., Desidera, S., Grundahl, F., and Lucatello, S. 2003, \aap, 408, 529
\bibitem[Hansen et al., 2007]{HN07} Hansen, B. et al. 2007, astro-ph/0701738
\bibitem[Hansen et al., 2004]{HN04} Hansen, B. et al. 2004, ApJS, 155, 551
\bibitem[Hansen et al., 2002]{HN02} Hansen, B. M. S.  et al. 2002, ApJL, 574, L155
\bibitem[Hansen, 1999]{HN99} Hansen, B. 1999, \apj, 520, 680
\bibitem[Hansen, 1998]{HN98} Hansen, B. M. S. 1998, Nature, 394, 860
\bibitem[Harris et al., 2006]{Har06} Harris, H. C. et al. 2006, \apj, 131, 571
\bibitem[Kilic et al., 2006]{Kilic06a} Kilic, M., von Hippel, T., Leggett, S. K., Winget, D. E. 2006, \apj, 646, 474
\bibitem[Koester \& Wolff, 2000]{Koester00} Koester, D. \& Wolff, B. 2000, \aap, 357, 587
\bibitem[Kowalski \& Saumon, 2006]{KS06} Kowalski, P. M., \& Saumon, D. 2006, ApJL, 651, L137
\bibitem[Richer et al., 2006]{RI06} Richer et al. 2006, Science, 313, 936
\bibitem[Richter et al., 2004]{RI04} Richer et al. 2004, \aj, 127, 2904
\bibitem[Richer et al., 2000]{RI00} Richer et al. 2000, \apj, 529, 318
\bibitem[Saumon \& Jacobson, 1999]{SJ} Saumon, D., \& Jacobson 1999, \apjl, 511, L107
\bibitem[Sirianni et al., 2005]{SIR05} Sirianni, M. et al. 2005, PASP, 117, 1049 
\end{thebibliography}
\end{document}